# COMPLETE NETWORK SECURITY PROTECTION FOR SME'SWITHIN LIMITED RESOURCES

Margie S. Todd[1], Syed (Shawon) M. Rahman[2]

[1]School of Business & Technology, Capella University, Minneapolis, MN, USA
[2]Associate Professor of Computer Science, University of Hawaii-Hilo, Hawaii, USA and
Part-time faculty, Capella University, Minneapolis, USA

## ABSTRACT

*The purpose of this paper is to present a comprehensive budget conscious security plan for smaller enterprises that lacksecurity guidelines.The authors believethis paper will assist users to write an individualized security plan. In addition to providing the top ten free or affordable tools get some sort of semblance of security implemented, the paper also provides best practices on the topics of Authentication, Authorization, Auditing, Firewall, Intrusion Detection & Monitoring, and Prevention. The methods employed have been implemented at Company XYZ referenced throughout.*

## 1.INTRODUCTION

Enterprise security has grown in complexity over the last decade. The goal of this publication is to show the proper methods of creating, implementing, and enforcing an enterprise information security plan. During the initial phase of the plan, the assessment, it is imperative that the shareholders of the enterprise identify the most valuable assets that need protection. The information technology department must be part of this determination so that strategies for network vulnerability mitigation can begin to emerge. Once there is a clear image of what it is that needs the most protection, then costing analysis can start. That is why identifying all possible vulnerabilities is critical so that if cost is an issue we can start prioritizing those assets in terms of what risks can the business afford versus what policies must be in place to avoid the exposure of the asset to the risk, by curbing employees' current behaviour.

Once the list of assets is solidified, begin conducting real time threat assessments keeping in mind that there are external (via Internet) as well as internal (employees) threats that must be considered. Each threat to our assets comes with its own vulnerability areas. Once the threats are identified we almost have to become criminal-minded to identify how much vulnerability accompanies each threat. Share the list with the shareholders and our management staff so that different perspectives are taken into consideration. Some of the vulnerabilities may be averted by creating policies for permissible network usage.

Write a comprehensive security plan for the enterprise and add it to our Business Continuity Plan and our Disaster Recovery Strategy. A comprehensive plan encompasses a clearly identified risk list with priorities and the impact of each security break in terms of how it affects the business. Supplement our security plan with a logging mechanism for all security events (real and perceived). If we create an event ticket tracking system it will encourage all employees to participate in the process thus reducing the time from reporting to containment. Be prepared to explain to the shareholders our rationale for our recommendations of one tool versus the other. The top ten affordable solutions are offered to minimize the initial cost of implementing some of the security risks by utilizing available tools from trustworthy sources to assist in software and hardware risk mitigation.

The next phase involves creating policies and procedures that support the security plan to be implemented. The policies must be polished and scrutinized so that we adhere to federal and state laws to ensure employee's rights are not violated; a feat somewhat easier to accomplish on the private sector. Once the policies are written and approved, then procedures can be written on carrying out the plan, the policies, and the consequences employees will face for failure to comply with the new secure plan. Once weare ready to implement this plan, and realizing that change is rather difficult, begin by conducting small trial runs and modify the plan and procedures accordingly. A proper implementation plan must be accompanied by a thorough training curriculum, although our goal is to devise a plan that is comprehensive yet simple enough that anyone in the corporation can easily follow it.

## 2. TOP 10 FREE OR NEARLY FREE SECURITY MEASURES TO IMPLEMENT RIGHT NOW

Many companies think of dollars signs when anyone mentions the word "security." If companies knew that the top 10 steps of basic security cost nothing at all – it costs nothing at all because it is something that, although in limited supply, still exists: common sense. Please consider the top 10 security audit practices provided by itsecurity.com as a checklist to secure our small business.

1)     **Know our equipment** – we do not know what weneed to secure and how unless we have a list of all our assets (including port numbers, ip addresses, printers, scanners, etc.) Name our assets so that they make sense to usand make auditing and inventorying easy. For instance, we are currently migrating from XP to Windows 7 or 8. One clever idea at our disposal is to name machines with the user's initials followed by the operating system and the year it was built. For example,consider naming the PC as MT713. Simply by looking at our network neighbourhoodwe can immediately tell what operating system they are on as well as the age of the device. In this case our machine is a Windows 7 machine and it was just replaced recent year. Why is this important? First of all we can keep track of the migration status, as well as,we can tell when something does not belong on that list.

2)     **Contain our threats by staying a step ahead** – identify all possible ways our network can be compromised and develop a plan to combat it, and in the worst case scenario make sure our crisis management plan includes a plan of action should our security is compromised.

3)     **Study our past so we will not make the same mistakes in the future** – studying past threats can help us to predict future ones. Use past mistakes as a training tool for our own users and our own support staff. Do not spend resources guarding the wrong assets. For instance, we get no walk-in traffic. All traffic is signed on and by appointment only. The computer room is not even in site of visitors. Why spend a ridiculous amount of money on cardkey access to the computer room if the threat is not there?

4)     **Prioritize our security concerns** – as mentioned above physical security is not a threat, however, is there a slight chance an employee has access to the computer and may take a spare monitor out of the room? Of course. However, the threat is so small that on the list of concerns this is probably near or at the bottom. Screen locks our servers so there is no chance users would sit at a server and use it. It is not a good practice to browse the Internet from a server as a threat can affect our main server and deploy through our entire network.

5) **Control access to our network** – there are free toolsavailable (a good source is CNET downloads) that allow us to design our own access control list. Make good use of groups and permissions. Any employee requiring access via Internet must pass one of our IT audits in order to connect. It must be done from an employer provided machine, must have the proper software and certificates installed and must abide by our strict computer and network usage policy. We do audit who logs on remotely and we record transactional logs of activity.

6) **Test our firewall** – there are some great freebies from our own software vendors that can test our firewalls for intrusions. Some utilities simply log the intrusions while other more advanced ones will react based on the threat. Reach out to our vendors.

7) **Give only what is necessary** – employees will be curious if we let them be. Some are not malicious they are just too curious for their own good. Only give them access to what is required. Do employ NTFS so access to folders is on a per-employee basis. We can take it a step further. We can designate an entire drive that is for manager content only. It simply just does not exist (it is not shared to authenticated users) nor is it mapped to everyone. One recommendation is to have ourmanagers' get into the good habit of locking their screens when they walk away from their desks. This minimizes the threat of someone sitting at a manager's desk and gaining access to some of the sensitive information.

8) **Backups, backups, backups** – this is so important for security purposes. There is a little dirty secret about backups though that we do not see discussed in any of the material we have read. SQL database backups are fantastic, however, when the software versions are upgraded those backup are no longer compatible with the software until and unless they are converted to the latest database version. Whenever there is a software upgrade those backups should be restored to a play database, converted, and then archived in the new format again. Why should this matter? A backup is only as good as being able to read and restore the data in it. An interesting suggestion for further research would be to make a study nationwide of how many companies do test their backups and of those, how many are actually able to use them right out of the box?

9) **All e-mail is evil** – teach users to suspect everyone. It is estimated that each day 55,000,000,000 spam e-mails circulate the cyber globe (itsecurity.com, 2007). Company XYZ had a sales manager who loved to read internet articles (for "research") and had the bad habit of sending hyperlinks with nothing else but his signature. He ended up sending a bad link to two employees and thank goodness for anti malware software both of them had the threat stopped. Needless to say I.T. had a conversation with him. Any technical person would have looked at the link and new it was not a valid hyperlink but that's just it – a technical person not an everyday Joe. Definitely it is worthy but the free version works well for a budget conscious company.

10) **The Tele-Worker Factor** –kids and our legged companions are a big threat to computer equipment (particularly those users who take laptops home). Stress the fact that the company issued laptop is not to be used for SpongeBob's revenge, nor is the warm top of the laptop a warm bed for a feline or canine companion. Enforce passwords on all portable devices. Yearly send out reminders to our home laptop users on proper care of their equipment.

# 3. PHYSICAL SECURITY MEASURES & RATIONALE

Data theft can occur by electronic means as previously mentioned; however the physical threats to information can arise from within the enterprise. As previously mentioned, XYZ Company's physical threats are defined as follows:

1) **Natural disasters** – although the office is located on the East Coast, far from water and in a traditionally non-tornado zone, last year brought management the reminder that mother nature listens to no one and follows no rules. With over 70 inches of snow, a tornado touchdown less than 10 miles away, and if that was not enough, an earthquake, disaster planning is a must. As such, the business is equipped with an A.D.T. alarm system so that if anything were to occur after the building is locked, senior management will be notified immediately. In addition, Company XYZ will immediately deploy the Crisis Management procedures (nowadays known as Disaster Recovery and Business Continuity Plan).

2) **Building Break In**– XYZ's office building is located inside a business park. It is safe to assume that non-business traffic is minimal, and it is the type of business where most (if not all) of the employees know who drives what type and model of car by simply walking out into the parking lot. Needless to say visitors stick out like a sore thumb. The building is closed to the public and there are signs everywhere that visitors must check in and check out at the front desk. As such, there is no reason for a company their size to invest in closed circuit video equipment, or key card access to the building. Our distribution industry (and the type of complex products they distribute) does not lend itself to walk-in business. All visitors must be announced and there is a clearly displayed door sign reading "NO SOLICITORS."

3) **Computer Break In-** these types can occur in the form of theft. One cannot prevent theft with certainty. Mobile and hand held devices can be forgotten, left behind, stolen, and even destroyed by accident. It is imperative that employees understand the role they play in securing these devices electronically and physically. It is so critical that we have made it a part of the Associate Handbook. Set up strong passwords; provide laptop carrying cases that are easy but secure and encourage employees to always carry their devices in the proper luggage. (Hint: emphasize the potential dangers of having radioactive devices in their pockets).

4) **Power Failure** – as a smaller company we may not be able to afford to have all computers connected to Uninterrupted Power Source (UPS) units. Rationally we have critical systems plugged into UPS systems (i.e. servers, firewall, phone system, T1s, packaging machines). We force all users to save their documents on employee shares on the network and we have auto save options set to ON for all PCs. When power failures occur we have the benefit to ascertain whether it is a long term power outage (by contacting the utility company) or if it is just a brown out. If it is a long term outage we then have the benefit of conducting a differential backup and shutting down through proper procedures. Our UPS boxes should sustain backup power for two hours which is ample time to conduct the necessary preparations in the event of a long term down time. Should the power outage occur after hours, ADT will contact all three individuals on the list and we will act accordingly. Weshould have access to the entire network remotely and can shut the system down completely.

In order to guarantee success it is advisable to consider the checklist provided here to determine if any of the risk factors deserve more than a once-a-year glance. These are (Ojdana&Watmore, p.5):

1. Can we detect [unauthorized] devices in real time including their physical location?
2. Can we automatically detect the root cause of an unscheduled disconnection of a critical device?
3. Can we enforce the compliance with prescribed security policy of a device connecting to the network?
4. Do we have the capability to deny an unauthorised or non-compliant device access to the network to prevent a potential threat risk?
5. Does our security plan take the physical infrastructure into account?
6. Is our network layout designed to be secure from intrusion?
7. Have all the policy and procedures for physical and connectivity access been documented for employees, contractors and service technicians?
8. Do our consultants and installation contractors have documented security policies and procedures in-line with mine?
9. Do our mission critical devices require higher security media such as fibre?
10. How would we be aware of a security breach in our physical network and shut it down?
11. Does our disaster recovery plan incorporate structured cabling requirements?
12. Does our network infrastructure use products with security features (i.e. [color] coded patch cords and modules, locking covers, termination mounts and/or pre-terminated fibre products)?
13. Do we have a real time schematic of the structured cabling installation including active ports?
14. Are all of our [organization's] floor-plans, hub-room drawings and port assignment documents up to date and in a secure location?

This list is not exhaustive but it provides the basics to develop a specific plan for our unique situation.

# 4. AUTHENTICATION

Another risk management factor is both user and process authentication (not to be confused with authorization discussed later on). Authentication is used to define two types of access: user access as well as devices or processes. (Oppenheimer,2010) Traditionally users have made up passwords for authentication based on one of the following:

1. Something they know – significant people's names, places, pets, favorite numbers
2. Something they have – entry keycards, bar-coded access cards, picture card, pin
3. Biometric – fingerprint, voice, retina scan, access code

All of these methods are discoverable by a multitude of techniques, some easier than others, including key logging, phishing, hijacking, data theft through accidental or purposeful discovery, and even social engineering. These risks have led to the recent deployment of two factor authentication.

Two factor authentication pioneers, SecurEnvoy and PasswordBank, have brought such services to cloud authentication. Penton Media Inc, presents an example on how these two partners collaborate to provide customers with such authentication. Users log onto one website with a traditional user defined parameter, and then the partnered companies provide the second factor utilizing SMS messaging to verify the identity. They list two benefits to having this convenience. First, it is managed purely on the cloud so it frees the worker to move about in the current global economy, second, the user is not handed any of the data to carry around increasing the risk of losing the information. With two factor authentication gaining momentum, this is actually a technology on the radar for Company XYZ to consider in the future for the outside sales team.

# 5. AUTHORIZATION

Authentication and authorization differ in that authentication occurs prior to accessing the secure network and authorization occurs after the network is accessed. Microsoft and most security professionals (if not all) recommend using the principle of least privilege. The Department of Defense Trusted Computer System Evaluation Criteria defines least privilege as "a principle that "requires that each subject in a system be granted the most restrictive set of privileges (or loyoust clearance) needed for the performance of authorized tasks. The application of this principle limits the damage that can result from accident, error, or unauthorized use." (Microsoft, 2006) In other words, give access exclusively to what the user absolutely needs to perform their duties, nothing else. Older software applications required PCs to have administrative privileges to run certain software. Nowadays that has changed by setting the controls based on policies rather than administrative access.

Nevertheless, users need to be curbed to not only run programs that are suitable for productivity without compromising the network, or access to files and folders that are within their domain and nothing further. At Company XYZthey have the practice of restricting manager's folders to only those with the Managers role. It is a shared drive so the drive only shows on manager's PCs so users do not even know the drive exists. It is worthwhile mentioning that although this is good enough to curtail user's access, it should be mandated that, as mentioned earlier, when manager's step away from their PCs for an extended time, they lock their screens or the user is setup with an auto locking desktop so that access is not granted by someone sitting at the manager's location.

Another area of concern for authorization and privileged accounts involves laptop users. Mobile workers may have the need to connect to multiple networks, go across continents, download data to portable drives, etc. Some IT departments do not apply the same rules to laptops as they do their infrastructures. A study conducted at the University of British Columbia found that all of the users in their study did not practice the least privilege principle and allowed all their laptops to have administrative privileges. (Motiee, Hawkey, Beznosv, 2010) According to their findings this alarming fact may be due in part to lack of training professionals on the benefits of using lower privileged accounts.

# 6. AUDITING

It is not enough to restrict usage, but it is also necessary to audit files and folders to ensure there are no loops that allow users to access restricted files. Before discussing the topic though, There's a need to clarify the fact that data stored on the Company XYZ network is considered intellectual company property. As such, anything stored on its server is company policy. This fact is part of the Computer and Network Usage policy at Company XYZ. When we design systems for the network it is recommended that we force all business software to automatically save the work to the employee's shared drive on the server. In order to balance the network load there are certain departments that are saved in separate servers. The dual purpose of this practice is to make auditing somewhat easier, but more importantly, so that the files are guaranteed to be backed up. When a PC is deployed to a user, he or she is trained on their responsibility for the tool. As such, it is emphasized that any personal property found on the server will be deleted immediately (no negotiations), and also, if the PC is rebuilt we use one image that does not allow for personal property to be restored.

There are some great tools available for auditing information and devices on the network. Regardless of what tool is used, practice human spot checks in strategic places where we suspect users "temporarily stash" temporary or junk information. In order to do spot checks we can use a utility named WinDirStat. The utility gives a visual depiction of drives as it colour codes file types. Company XYZ has used it in the past when a rogue log file was filling up a hard drive form a RAID array. It was pretty handy as the rogue file lit up like a Christmas tree. There is a third tool called "experience." After so many years inspecting the network,we willdevelop a mental picture of what should go where and files that do not belong make themselves highly visible when auditing is the norm. Although automated for the most part, there is still a need for human inspection.

When it comes to hardware, Company XYZutilizes Spiceworks as both an auditing, inventory tool, as well as the IT ticket log that allows users to submit and keep track of IT requests integrated with our Intranet portal. If we do not have an Intranet (internal website available to internal network users only) it is recommended that we set one up. It may be free to setup as long as we install the Internet Information Services (IIS) role that is often with some versions of the Microsoft operating system software. There are plenty of free tutorials on the Internet on how to create it using the Microsoft tools we may already have in place.

## 7. IMPLEMENT A FIREWALL

This is the most important element of network security and perhaps the least understood by senior management during the risk mitigation evaluation phase. Most users do not understand that firewalls are both hardware and software based. Some professionals seem to be of the opinion that a software firewall is good enough, some others believe a good compliment of both works best. For a business of any size that is exposed to the entire world through the Internet should invest in a robust firewall solution. Please examine how cyberspace has changed in recent times by studying the following facts provided by FireEye (2013):

1. Malware has been identified as originating from 184 distinct countries around the globe.
2. 89% of the tools utilized in this type of activity originated in China
3. Targets of these attacks are in specific industries (healthcare, tech companies, utilities, manufacturing, logistics, transportation, etc)
4. U.S. is still the top target of the attacks (taking 66% of all the hits)
5. Technology is constantly evolving – social networking, data hijacking
6. Attacks vary globally
7. It is a constant globally moving target

The list is not all inclusive; however, it is enough to make any security professional take a deep look at the methods being employed to thwart these efforts. Static and single targeted approaches to security are not adequate anymore. Other important evolving technologies are worth adding to the list: remote access, denial of service attacks, botnets, war driving, mobile infection, etc. The best approach is to deploy a firewall that "learns" behaviour and patterns and can adapt. There are vendors out there that provide such a tool.

## 8. INTRUSION DETECTION & PREVENTION MONITORING

Although some of the detection capabilities are handled by the ASA device aforementioned, there is still a need to perform preventive monitoring. Sometimes the information leaving or transacting within the network is far more malicious than the information arriving from the Internet. There is physical detection that can occur on a sporadic basis. Oppenheimer asserts that some of the false alarms that the automated monitoring creates are more annoying than helpful (Oppenheimer, 2010) As a smaller company we have to pick which battles we engage and which ones we hand off to a device. It is more critical in our particular environment that we protect the network from within. Company XYZ has over 360 years in tribal knowledge and business intelligence due to the tenure of their employees; that needs to be preserved and closely monitored. Realize that our efforts are mostly preventive.

A comprehensive preventive plan must take into account the four major areas of the detection and prevention technologies (Scarfone&Mell, 2007):

1) **Network Based** – this is accomplished by using tools to analyse internal network traffic, such as Wireshark™, which tracks every bit of data moving throughout the network. It can be deployed in single and promiscuous mode. The latter monitors all traffic going through the network as opposed to one system.

2) **Wireless** – if we decide that our corporate policy does not allow wireless connections to the network so the problems normally encountered with this type of threat are not a "battle" we care to engage as mentioned above.

3) **Network Behaviour Analysis** – this is accomplished by analysing traffic arriving at our firewall for problems such as policy violations, denial of service attacks, malware, war driving, etc. Partner up with our Internet Service Provider (ISP) who provides we with a service that allows us to log onto ourconnection (whatever that may be – some legacy products are not as advanced) to monitor usage at any given time. The ISP monitors the traffic on their own as well and they notify us if our average consumption varies beyond certain parameters we set in place when they installed their line.

4) **Host-Based** – this is accomplished by monitoring a defined host or service from a host for problems as well.

All these tools are required in order to provide a comprehensive plan to monitor the network for intruders. It would be irresponsible if factors such as performance, service expectations, technical support, cost, and training are not taken into consideration into the plan.

## 9.INFORMATION SECURITY POLICIES AND PROCEDURES

The biggest hurdle is creating all encompassing security policies and procedures that will address every situation imaginable. Realizing that it is impossible to be 100% safe all the time, the task will actually become somewhat easier. Start with the most obvious and build on it. We do not want to be the target of every attack before developing a comprehensive plan but we have to start somewhere. Both Microsoft (MCS, 2007) and the C.I.A. have templates of policies that can be adopted and modified to fit each enterprise. The policies will in turn dictate the procedures as these are unique to each company. Our security policies are clearly stated in our Associate Employment Handbook. The procedures are then governed and reviewed by our ISO certified quality program.

Information security policies must be first discussed and agreed to by the entire management staff. The policies set forth must be in tune with the current industry. For instance, examine the following social engineering policy. Even though we say:

"XYZ Company, Inc. is increasingly exploring how online discourse through Social Media can empower XYZ Company, Inc. as an industry leader. It is very much in the XYZ Company, Inc interest to be aware of and participate in this sphere of information, interaction and idea exchange.

The same principles and guidelines that apply to XYZ Company, Inc associates' activities in general also apply to employees' Social Media activities via Facebook, MySpace, LinkedIn, YouTube, Plaxo, Twitter, etc.  Social media describes the online technologies and practices that people use to share opinions, insights, experiences, and perspectives. Social media can take many different forms, including text, images, audio, and video. Social Media sites typically use technologies such as websites, blogs, message boards, podcasts, wikis, and blogs to allow users to interact" (XYZ Company, 2012).

On the same policy we explicitly state:

"Personal Social Media activities must not take place during work hours or using XYZ Company, Inc equipment. Refer to the XYZ Company, Inc Computer Policy" (XYZ Company, 2012).

They are not Neanderthals, they obviously understand that social media is here to stay. However, Company XYZ's industry is not well served by social interaction with the public. They provide sealing solutions to complex environmental and chemical problems. In addition to the information not being highly exciting and complicated, there is no value to "liking" XYZ Company on Facebook as it brings no value to the general consumer. If theywere in the retail business perhaps then it would benefit us from participating in such activities. There is an inherent danger in social engineering and the spread of worms, trojans, and viruses through participating n them with friends. Why take the risk if the reward is zero?

## 10.SECURITY PLAN IMPLEMENTATION

It is great to have a plan drafted on paper but it is worth its weight in gold if it is not properly implemented. The biggest hurdle to get over is answering the two questions that burn in everyone's mind when the issue of security policy implementation comes up: Why and How much? It is bothersome how in a post 9/11 world there are still people that ask the why question. In a world riddled with wars, conflicts, never before seen weather phenomenon, an overactive sun, aging equipment, etcwe must not question if we will have to deal with a major security event in our lifetime but rather when it will occur. Have we ever considered what would happen if a pay check did not come during the next pay period? How about for the next six pay periods? Regardless of who is asked, the answer is probably the same: We don't know. What if there was a plan developed so that mitigation was in place ahead of time should the situation arrive? Think of equating protecting the enterprise to protecting a pay check (in reality that is what it is all about). How much will it cost? Wewere presented information earlier that showed how a basic security plan does not have to be expensive. Most of it can be done with existing resources and relies heavily on common sense and education.

## 11.TRAINING

As employees continue to be the highest information security threat, wecan make it a departmental goal to conduct a set number of training sessions per year. Some training is mandatory (such as security) and some is voluntary. The year of 2001 was very defining for many companies as up until then a good portion of the population lived in denial. The September 11th attacks brought about major changes in the industry. Training and awareness became the focal point and continues to be today. Back then Company XYZ started developing training guides for the company that encompassed all areas of the business. They had some training in place but realized the curriculum and monitoring was poor. I.T. began putting together a curriculum that continues in existence to this day and gets reviewed on a constant basis. Every other year has become a training year for every department.

During the year of 2012Company XYZ's I.T. department conducted 100 employee training sessions – they spent almost half a business year training as well as dealing with regular IT issues. The year of 2011 became the train-the-trainer year. They partnered with a local IT training school (New Horizons Computer Learning Centers) with a special pricing program that allowed any of their IT employees to attend any online classes any time of the year. The amount of confidence an employee gains from proper training is priceless. The work gets done more efficiently and deployment becomes a very smooth process. It also brings light to the new security threats and countermeasures based on all the new software changes and the improvement in monitoring that occur with upgrades. It also forces the company to review existing processes to enhance them with newerr updated information.

## 12. PREVENTION

Taking education to the next step is another way to combat threats to the enterprise network. The world may not want to admit it but we are clear in the middle of a cyber-war. Countries are getting more and more sophisticated with their attacks, there is blatant web espionage, home grown cyber terrorism and the threats and creativity is not showing signs of a slowdown. Eternal vigilance is the only antidote to enterprise risk. I.T. professionals need to use the Internet as a research tool to stay ahead of the emerging threats as well as get educated on how to learn to monitor the network and scope out those new threats. There are several ways to do that:

1. **Research online** – subscribe to websites that monitor and publish emerging threats. (Example: http://www.capella.edu/blogs/iascommunity/ ). This is a good place for new research.
2. **Trade publications** – read as much as we can about predictive articles. Example: "The biggest cybersecurity threats of 2013" (Teller, 2012)
3. **Professional association websites** – SANS Institute
4. **Live risk monitoring services** – websites dedicated to the constant round the clock monitoring and reporting of newly discovered threats (Example: https://isc.sans.edu/ )

These are some of the self-paced educational opportunities to keep our IT staff abreast of the latest emerging threats to the network environment. An approach that may work for our business is to create educational incentives built into an employee's salary package. The standard is set so that the more we demonstrate our learning growth (tracked through Success Factors™) the higher the salary increase will be at performance appraisal time. It gives them an incentive to learn, and the return on investment for the company is achieved through performance and productivity gains from the newly found and implemented risk mitigation methods.

## 13. CONCLUSIONS

The overall goal of this article was to create a budget conscious security plan after a thorough analysis of the enterprise. Readers will be able to draft, organize and create a comprehensive security plan by following the recommendations presented. The plan will be comprised of all the necessary components of a thorough enterprise analysis such as: preliminary security assessment, security requirements, security plan, security plan policies, and security procedures. Basic and affordable security monitoring recommendations are also presented to get an enterprise headed in the proper direction to create a culture of security minded employees to survive current and emerging network security threats.

**Authors Bio:**

**Margie S. Todd**is the Chief Technology Officer for an industrial distributor in the greater Philadelphia region. She has a B.A. degree from the University of Connecticut. Margie is currently pursuing a Ph.D. in Information Security and Assurance from Capella University. She has expertise in software and hardware analysis as well as over 18 years of experience in the field of information technology. She serves on Board of Advisors, writes professional product reviews, and researches emerging technologies. Her personal interests are reading, competitive billiards, and volunteer work for police charities 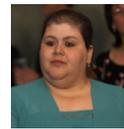

**Dr.Syed (Shawon) M. Rahman** is an Associate professor in the Department of Computer Science and Engineering at the University of Hawaii-Hilo and a part-time faculty of information Technology, information assurance and security program at the Capella University. Dr. Rahman's research interests include software engineering education, data visualization, information assurance and security, web accessibility, and software testing and quality assurance. He has published more than 90 peer-reviewed papers. He is a member of many professional organizations including ACM, ASEE, ASQ, IEEE, and UPE.. 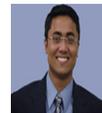